\documentclass[twocolumn,showpacs,preprintnumbers,amsmath,amssymb]{revtex4}
\usepackage{graphicx}
\usepackage{dcolumn}
\usepackage{bm}

\begin{document}

\title{Circularly polarized electroluminescence
in spin-LED structures}

\author{Z. G. Yu, W. H. Lau, and M. E. Flatt\'e}

\affiliation{Department of Physics and Astronomy, University of Iowa,
Iowa City, Iowa 52242}
\date{\today}

\begin{abstract}
We calculate circularly polarized luminescence emitted  parallel
(vertical emission) and perpendicular (edge emission) to the
growth direction from a quantum well in a spin light-emitting diode (spin-LED)
when either the holes or electrons are spin polarized. It is essential to account for the {\it
orbital} coherence of the spin-polarized holes when they are
captured in the quantum well
 to understand recent experiments demonstrating polarized edge
emission from hole spin injection. The calculations  explain many features of the
circular polarizations of edge and vertically emitted luminescence for spin polarized hole injection from Mn-doped
ferromagnetic semiconductors, and for
spin-polarized electron injection from II-VI dilute magnetic semiconductors.

\end{abstract}
\phantom{.}

\maketitle

\begin{figure}
\includegraphics[width=8cm]{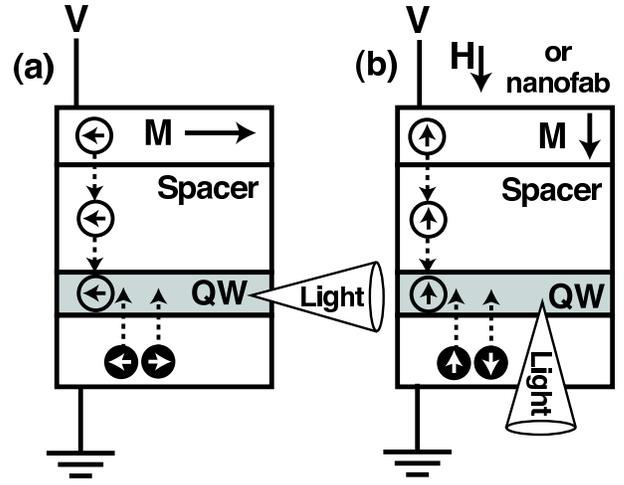}
\caption{Schematic showing the spin-LED geometries for edge
emission (a) and vertical emission (b). Shape anisotropy tends to force situation (a) without a large vertical magnetic field or nanofabrication.}\label{diagram}
\end{figure}

One of the most widely-used semiconductor spintronic devices
\cite{wolf,spinbook} is the spin LED\cite{molenkamp,ohno}. In a spin LED circularly polarized
light is emitted after the recombination of spin-polarized
carriers that are electrically injected into a semiconductor
heterostructure. This device is commonly used to measure the
spin injection efficiency into
materials\cite{molenkamp,ohno,jonker,ploog,crowell,young,hanbicki}. It is
therefore vital to quantitatively understand the circular
polarization of luminescence ($P_\ell$) in order to
 accurately determine
the spin injection efficiency or local spin polarization in a
semiconductor. There are two typical spin-LED geometries: the
light can come out the edge of the device, or vertically out the
top or bottom. As shown in Fig.~\ref{diagram}(a) the edge-emitting structures
are designed to inject carriers with spin perpendicular to the
growth direction, while in Fig.~\ref{diagram}(b) the vertical-emitting devices inject carriers with parallel spin. The selection rules for vertical
emission are reasonably straightforward -- that is not the case
for edge emission. A theory based on simple selection rule
arguments would suggest zero circular polarization of the
edge emission for both electron- and hole-spin injection
because of a large energy splitting between heavy- and light-hole
states in the typical recombination region, a quantum well (QW).
Experiments, however, have demonstrated  that although the
$P_\ell$ in Fig.~\ref{diagram}(a)  is much weaker
than in Fig.~\ref{diagram}(b), hole-spin injection can lead to a sizable $\sim
1$\% circular polarization of edge emission\cite{young}, although of opposite sign. Meanwhile, reports of edge $P_\ell$ from electron-spin injection conflict; in a Zener tunneling diode\cite{youngsst} it equals the vertical emission $P_\ell$, whereas for a ZnMnSe injector the edge $P_\ell$ is negligible\cite{molenkamp2}. Although the selection rules appear more subtle, the device geometry of Fig.~\ref{diagram}(a) 
has important advantages over Fig.~\ref{diagram}(b).  Large magnetic fields
 or sophisticated fabrication 
are generally required to orient the spin out-of-plane [Fig.~\ref{diagram}(b)]\cite{molenkamp,jonker,ploog,crowell,young,hanbicki}. Hence a quantitative
understanding of the origin of polarized edge emission can
lead to simpler devices which detect accurately the spin
polarization of carriers.

In this letter we resolve many of these discrepancies between the experiments and the theory,
by describing circularly polarized edge and vertical
emission from spin LEDs for both hole- and electron-spin injection
within a unified framework that includes the potential for orbital coherence within the QW. The $P_\ell$ depends on
several processes: injection from the magnetic semiconductor into the spacer, transport through the spacer, transport from the spacer to the QW, and the recombination dynamics within the QW. Spin polarization decay lengths within the spacer are seen experimentally to differ significantly depending on the orientation of the hole spins --- this is most likely due to anisotropic spin relaxation parallel and perpendicular to carrier motion\cite{shayegan-winkler}. Our principal interest is in the question of how the QW luminescence indicates the local spin polarization. Thus we focus on $P_\ell$ in Fig.~\ref{diagram}(a) and (b) for the smallest spacer thickness.
 A key new ingredient in our theory is that
the {\it orbital} coherence of carriers can be retained when they move
from the spacer to the QW. Our calculations  can explain many features of the circular polarizations of edge and
vertical emission in spin LED's for hole spin injection from Mn-doped ferromagnetic semiconductors. Due to their orbital degree of freedom,  spin-polarized holes give rise to a
much stronger edge $P_\ell$ than
spin-polarized electrons.

{\it Edge emission for hole-spin injection [Fig.~\ref{diagram}(a)].} The
spin LED structure in Ref. \cite{ohno} consists of a $p$-type
ferromagnetic semiconductor (GaMnAs), a spacer (GaAs), and
a QW (In$_{0.13}$Ga$_{0.87}$As). When spin-polarized holes are
injected from the ferromagnetic semiconductor into the spacer,
 they predominately occupy the heavy-hole states of  bulk GaAs. As the orbital quantization axis is parallel to the spin for heavy holes, the states occupied must have a preferred orbital orientation as well, for
$|{\bf k} \uparrow(\downarrow) \rangle = \frac{1}{\sqrt{V}}|v_{\uparrow
(\downarrow)}\rangle e^{i {\bf k}\cdot {\bf r}}$, where $V$ is the volume, and
\begin{equation}
|v_{\uparrow}\rangle=\frac{1}{\sqrt{2}}(Y+iZ)
|\uparrow_x\rangle,~~
|v_{\downarrow}\rangle=\frac{1}{\sqrt{2}}(Y-iZ) |\downarrow_x\rangle.\label{def-HH}
\end{equation}
Thus for holes, spin polarization selects an orbital wavefunction. The density matrix of holes in the spacer,
\begin{equation}
\rho^h({\bf k})=f^h_{\uparrow}(E_{\bf k})|{\bf k}\uparrow \rangle \langle {\bf
k}\uparrow|
+f^h_{\downarrow}(E_{\bf k})|{\bf k}\downarrow \rangle \langle
{\bf k}\downarrow|.
\end{equation}
$f^h_{\uparrow(\downarrow)}(E)=[1+\exp\{(\mu^h_{\uparrow(\downarrow)}-E)/k_BT\}]^{-1}$
is the Fermi distribution for up-spin (down-spin) holes and
$\mu^h_{\uparrow(\downarrow)}$ is the up-spin (down-spin) Fermi energy. The spin
polarization of the (non-degenerate) spacer carrier density,
\begin{equation}
P_{d,s} = { {\rm e}^{\Delta\mu^h/k_BT} - 1\over {\rm e}^{\Delta\mu^h/k_BT} +1},
\end{equation}
depends on $\Delta \mu^h = \mu^h_{\uparrow} -\mu^h_{\downarrow}$.

A 14-band envelope-function  ${\bf K}\cdot{\bf p}$ calculation generates the
electronic structure of the QW. Such a calculation accurately predicts the spin splitting in a variety of semiconductor heterostructures\cite{kdp,spinbook}. Each
eigenstate in the QW is labelled by the momentum {\bf K} and an index $L$ for the other quantum numbers.

When the spin-polarized holes move into the QW from the spacer, they would keep
their heavy-hole (both orbital and spin) character, $|v_{\uparrow(\downarrow)}\rangle$, if they could.
 $|v_{\uparrow(\downarrow)}\rangle$ is not an eigenstate in the QW so
the
holes
will relax into QW eigenstates. We describe this relaxation process  beginning with
a density matrix for holes in
the QW,
\begin{eqnarray}
\rho^{h}_{LL'}({\bf K},{\bf K'})&=&\phi_{LL'}({\bf K},{\bf K'}) |L{\bf K}\rangle \langle
L'{\bf K'}|,\\
\phi_{LL'}({\bf K},{\bf K'})&=&\phi^{\uparrow}_{LL'}({\bf
K},{\bf K'})\langle
L{\bf K}|v_{\uparrow}
\rangle
\langle
v_{\uparrow}|L'{\bf K'} \rangle\nonumber\\
&+&\phi^{\downarrow}_{LL'}({\bf K},{\bf K'})\langle L{\bf K}|v_{\downarrow}
\rangle \langle
v_{\downarrow}|L'{\bf K'}\rangle.\label{dmatrix-def}
\end{eqnarray}
This description is central to this paper. It takes into account the wave
function overlap between the spacer's (bulk) heavy-hole states
$|v_{\uparrow(\downarrow)}\rangle$ and the QW eigenstates $|L{\bf
K}\rangle$, as well as the carrier distribution function imposed by the spacer (at  steady state the spin-dependent electrochemical potentials are continuous
across the interface between the spacer and the QW). 

 If the elastic mean free path of the holes were infinite, only
matrix elements that connect states with the same energy in the
density matrix would survive, and
\begin{equation}
\phi^{\uparrow(\downarrow)}_{LL'}({\bf K},{\bf K'})=f^h_{\uparrow(\downarrow)}(E_{L{\bf
K}})\delta(E_{L{\bf
K}}-E_{L'{\bf K'}}).
\end{equation}
The actual finite scattering rate broadens
the delta function in the above equation --- here the functional form is assumed to be Gaussian, 
\begin{eqnarray}
\phi^{\uparrow(\downarrow)}_{LL'}({\bf K},{\bf K'})&=&\pi^{-1/2}\alpha^{-1}f^h_{\uparrow(\downarrow)}[(E_{L{\bf K}}+E_{L'{\bf K'}})/2]\nonumber\\
&&\times \exp[-
(E_{L{\bf K}}-E_{L'{\bf K'}})^2/ \alpha^2],
\end{eqnarray}
where the parameter $\alpha$ characterizes the scattering rate (assumed equal for all states).
In the numerical
calculations presented here, $\alpha$ is chosen to be 2 meV,
corresponding to a $\sim 0.5$~ps scattering time. 
The numerical results are not sensitive to the precise value of $\alpha$
so long as $\alpha \gtrsim 0.1$~meV.
The electrons, by contrast, are unpolarized and in equilibrium. Their density matrix,
\begin{equation}
\rho^{e}_{\tilde{L}\tilde{L'}}({\bf
K},{\bf K'})=g_{\tilde{L}\tilde{L'}}({\bf K},{\bf K'})
|\tilde{L}{\bf K}\rangle \langle \tilde{L'}{\bf K'}|.
\end{equation}
$\tilde{L}$ is the conduction band index,
$g_{\tilde{L}\tilde{L'}}({\bf K},{\bf K'})=f^e(E_{\tilde{L}{\bf
K}})\delta_{\tilde{L}\tilde{L'}}\delta_{{\bf K}{\bf K'}}$, and
$f^e(E)=[1+\exp (E-\mu^e)]^{-1}$.

The circularly polarized luminescence,
\begin{eqnarray}
I_{\pm}(\omega) =&& \sum_{\tilde{L}\tilde{L'}LL',{\bf
K},{\bf K'}}g_{\tilde{L}\tilde{L'}}({\bf K},{\bf K'}){\bf
d}^{\pm}_{\tilde{L'}L'}({\bf K'}) \phi_{L'L}({\bf K'},{\bf K}) \nonumber\\
\times
{\bf
d}^{\pm}_{L\tilde{L}}\kern-5pt\ ^*({\bf K})&&\delta(\omega-(E_{\tilde{L}{\bf
K}}+E_{\tilde{L'}{\bf K'}}-E_{L{\bf K}}-E_{L'{\bf K'}})/2),\label{intensity}
\end{eqnarray}
where  $+(-)$ represents right- (left-) circular polarization of
the light.  The matrix elements of the dipole operator
\begin{equation}
{\bf d}^{\pm}_{\tilde{L}L}({\bf K}) =\langle
\tilde{L}{\bf K}|\frac{1}{\sqrt{2}}({\bf e}_y
\mp i{\bf e}_z)\cdot {\bf P} |L{\bf K}\rangle.\label{dipole-op}
\end{equation}
As $g_{\tilde{L}\tilde{L'}}({\bf K},{\bf K'})$ is diagonal in the momentum and band index, the ${\bf K'}$ and $\tilde{L'}$ sums in Eq.~(\ref{intensity}) can be done trivially by replacing ${\bf K'}$ and $\tilde{L'}$ by ${\bf K}$ and $\tilde{L}$.
The circular polarization of luminescence is defined as
$P_\ell =(I_+ -I_-)/(I_+ + I_-)$. 
The spin polarization of density in the QW can be calculated via
\begin{equation}
P_d \equiv
\frac{n_{\uparrow}-n_{\downarrow}}{n_{\uparrow}+n_{\downarrow}} =\langle
\sigma_x\rangle
=\sum_{LL'{\bf
K},{\bf K'}}\frac{Tr[\sigma_x
\rho_{LL'}({\bf K},{\bf K'})]}{Tr \rho_{LL'}({\bf K},{\bf K'})}.\label{dens-def}
\end{equation}
Equations~(\ref{def-HH})-(\ref{dens-def}) permit orbital coherence to be partially
maintained in our calculation -- an effect necessary for
explaining the edge $P_\ell$ for hole spin LEDs.

\begin{figure}
\vspace{10pt}
\includegraphics[width=8cm]{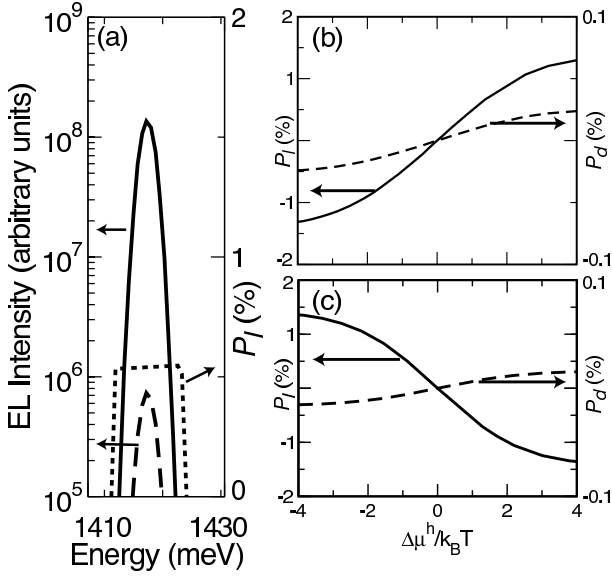}
\caption{(a) Total edge emission intensity, $I_+ + I_-$ (thicker solid
line), the intensity difference between right-circularly and
left-circularly polarized emission, $I_+ - I_-$ (dashed line) and
luminescence polarization $P_\ell$ (dotted line) for heavy-hole spin injection into a 100
\AA~ In$_{0.13}$Ga$_{0.87}$As
 QW with hole
density $10^{15}$ cm$^{-3}$ and $P_{d,s} = 52$\% ($\Delta \mu^h =0.57$~meV).
(b) energy-integrated edge-emission polarization
(solid line)
and
$P_d$ (dashed line) versus $\Delta\mu^h$. (c) same as (b) but for light-hole spin injection. The temperature is 6K.
A 3 meV Gaussian linewidth smoothes the luminescence spectrum. }\label{hole-edge}
\end{figure}

\begin{figure}
\includegraphics[width=8cm]{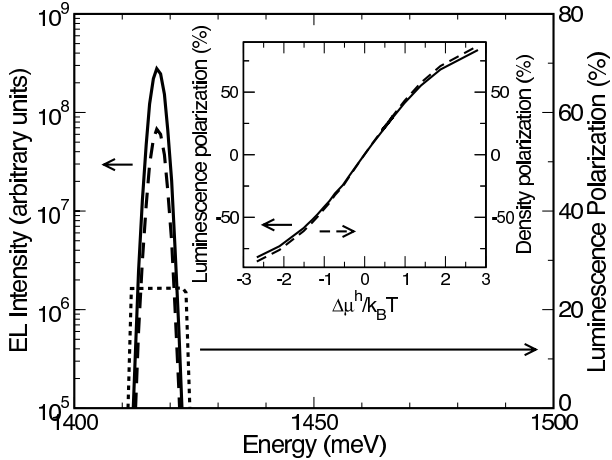}
\caption{Vertical emission from the system in Fig.~\ref{hole-edge}(ab) with the same parameters except $P_{d,s} = 27$\% ($\Delta \mu^h =0.28$~meV).}\label{hole-vert}
\end{figure}

Figure~\ref{hole-edge}(a) shows spectra of the total electroluminescence and the edge $P_\ell$ for heavy-hole spin injection.
The edge emission is circularly polarized
and the polarization can easily reach 1\%. The maximum circular
polarization is 1.4\% when fully spin-polarized heavy holes are injected into
the spacer ($\Delta\mu^h \to \infty$). Fig.~\ref{hole-edge}(b)  displays the
circular polarization of the energy-integrated luminescence and
the spin polarization of density as a function of $\Delta\mu^h$.
$P_\ell$ increases monotonically with $P_{d,s}$,
indicating that the edge $P_\ell$ can be
used to accurately  measure the spin injection efficiency from the magnetic contact into the spacer.   The energy splitting of the heavy holes and light holes in the QW does suppress $P_d$ and $P_\ell$ relative to $P_{d,s}$.  The $P_\ell$, however, still greatly exceeds $P_d$ due to the {\it remnant carrier orbital coherence} in the QW. Figure~\ref{hole-edge}(c) shows the same quantities as Fig.~\ref{hole-edge}(b), but for light-hole spin injection. The different orbital character of the holes in the spacer leads to a sign reversal of $P_\ell$.

{\it Vertical emission for hole spin injection [Fig.~\ref{diagram}(b)].} The magnetization of the ferromagnetic semiconductor
is now along the $z$-axis. $P_\ell$ and $P_d$
can be computed as before by replacing $|v_{\uparrow}\rangle$ and
$|v_{\downarrow}\rangle$ in Eq.~(\ref{def-HH}) by
$|v_{\uparrow}\rangle=\frac{1}{\sqrt{2}}(X+iY)|\uparrow_z\rangle$ and
$|v_{\downarrow}\rangle=\frac{1}{\sqrt{2}}(X-iY)|\downarrow_z\rangle$,
${\bf d}^{\pm}_{\tilde{L}L}({\bf K})$ in Eq. ~(\ref{dipole-op}) by $ {\bf
d}^{\pm}_{\tilde{L}L}({\bf K}) =\langle \tilde{L}{\bf
K}|\frac{1}{\sqrt{2}}({\bf e}_x \mp i{\bf e}_y)\cdot {\bf P}
|L{\bf K}\rangle $, and $\sigma_x$ in Eq.~(\ref{dens-def}) by $\sigma_z$. 
The spectra of luminescence along the $z$-axis and $P_\ell$ are shown in Fig.~\ref{hole-vert}. This
configuration leads to a much stronger circular polarization for heavy hole spin injection than
the edge-emission configuration, which is consistent with  the
experimentally observed 14-fold ratio of the circular
polarization between the two configurations\cite{young}. Light hole spin injection creates a negligible signal for vertical emission.

The spin and orbital contribution to the emission polarization can be analyzed as a function of momentum ${\bf K}$. The time-reversal symmetry of the system implies that
a state $|L{\bf K}\rangle$ and its time-reversal state $T|L{\bf K}\rangle \to
|{\cal L}{\bf -K}\rangle$, are related by $E_{L{\bf K}}=E_{{\cal
L}{\bf -K}}$. These two states have opposite (pseudo)spin
orientations. Noting that
$d^{\pm}_{\tilde{L}L}({\bf K})=(d^{\mp}_{\tilde{\cal L}{\cal L}}({\bf -K}))^*$,
and the electrons
are not spin-polarized, the polarization of light
for
each momentum ${\bf K}$ is
\begin{equation}
\frac{\phi_{LL'}({\bf K},{\bf K})-\phi_{\cal LL'}(-{\bf
K},-{\bf K})}{\phi_{LL'}({\bf
K},{\bf K})+\phi_{\cal LL'}(-{\bf K},-{\bf K})}={\cal D}_{LL'}({\bf K})W_{LL'}({\bf K}),
\end{equation}
where 
\begin{eqnarray}
{\cal D}_{LL'}({\bf K})&=&\frac{f^h_{\uparrow}[(E_{L{\bf K}}+E_{L'{\bf K}})/2]
-f^h_{\downarrow}[(E_{L{\bf K}}+E_{L'{\bf K}})/2]}{f^h_{\uparrow}[(E_{L{\bf K}}+E_{L'{\bf K}})/2]
+f^h_{\downarrow}[(E_{L{\bf K}}+E_{L'{\bf K}})/2]}\nonumber\\
W^h_{LL'}({\bf K})&=&\frac{\langle L{\bf
K}|v_{\uparrow}\rangle
\langle v_{\uparrow}|L'{\bf K}\rangle-\langle L{\bf K}|v_{\downarrow}\rangle
\langle
v_{\downarrow}|L'{\bf K}\rangle}{\langle L{\bf
K}|v_{\uparrow}\rangle
\langle v_{\uparrow}|L'{\bf K}\rangle+\langle L{\bf K}|v_{\downarrow}\rangle
\langle v_{\downarrow}|L'{\bf K}\rangle}.
\end{eqnarray}
Thus the $P_\ell$ is determined not only by $\Delta\mu^h$ but
also the characteristics of the wave functions in the spacer and in
the QW. For vertical emission the character of the QW eigenstates at the top
of the valence band is very similar to the polarized
heavy-hole states in the spacer, so $W^h_{LL'}$ is close to $\pm\delta_{LL'}$
and $P_\ell \approx P_{d,s}$. For edge emission the process is more complex. First, holes are spin
polarized in the spacer (characterized by ${\cal D}_{LL'}$). This spin polarization in the spacer is converted to a predominately
orbital coherence in the QW (off-diagonal components of $W^h_{LL'}$), which produces circularly polarized
luminescence.

\begin{figure}
\includegraphics[width=8cm]{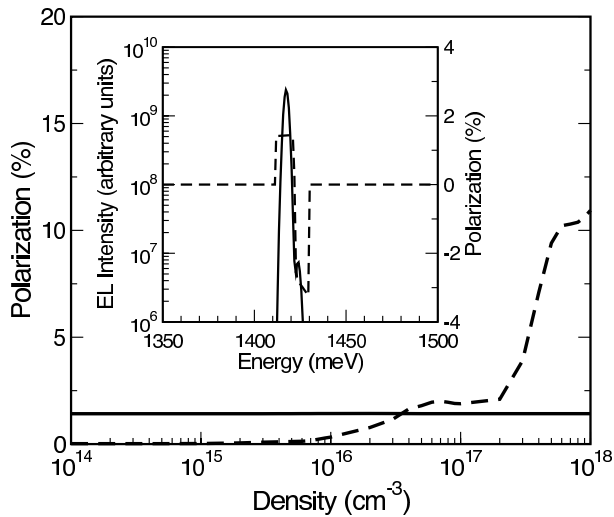}
\caption{Maximal $P_\ell$ (solid line) and $P_d$ (dashed
line) in the QW as a
function of hole density for in-plane emission at $T=6$K.
The inset illustrates the total electroluminescence (solid line) and
$P_\ell$ with hole density $10^{16}$ cm$^{-3}$.
}\label{hole-high}
\end{figure}

For larger hole densities in the QW the carriers will
occupy QW eigenstates with higher energies and different
character. Figure~\ref{hole-high}(inset) shows the edge emission and
the circular polarization for fully polarized ($\Delta\mu^h
\to \infty$) holes  with density 10$^{16}$ cm$^{-3}$. A new peak emerges
at a higher energy in the luminescence spectrum. The circular
polarization at the high-energy peak has the opposite sign as that of 
the low-energy peak, suggesting that the injected holes may begin to
enter QW states with light-hole character. The energy-integrated
circular polarization over the low-energy peak region, however,
is independent of the carrier density over the entire
experimentally accessible range, as illustrated in Fig.~\ref{hole-high}. This agrees 
with the measured insensitivity of $P_\ell$
 to the current. Figure~\ref{hole-high} also shows
that $P_d$ increases with the total
hole density in the QW (states with
light-hole characteristics can, with the heavy holes, form a state with spin along the
$x$-axis).

{\it Emission from electron spin injection.} The
wavefunctions of injected electrons in the spacer region are
$|k\uparrow(\downarrow)\rangle =
\frac{1}{\sqrt{V}}|c_{\uparrow(\downarrow)}\rangle e^{i{\bf k}\cdot {\bf
r}}$, $|c_{\uparrow(\downarrow}\rangle=|S_{\uparrow_x (\downarrow_x)}\rangle$
for the in-plane configuration and
$|c_{\uparrow(\downarrow)}\rangle=|S_{\uparrow_z(\downarrow_z)}\rangle$ for
the $z$-axis configuration. The density matrix of electrons is
constructed as in Eq.~(\ref{dmatrix-def}) with $|v_{\uparrow(\downarrow)}\rangle $
substituted by $|c_{\uparrow(\downarrow)}\rangle$. The holes are
unpolarized and only the diagonal elements in the hole density
matrix have nonzero values. As the electrons have no orbital degree of freedom comparable to that of the heavy holes, the expectations for edge emission should be very different.

\begin{figure}
\vspace{10pt}
\includegraphics[width=8cm]{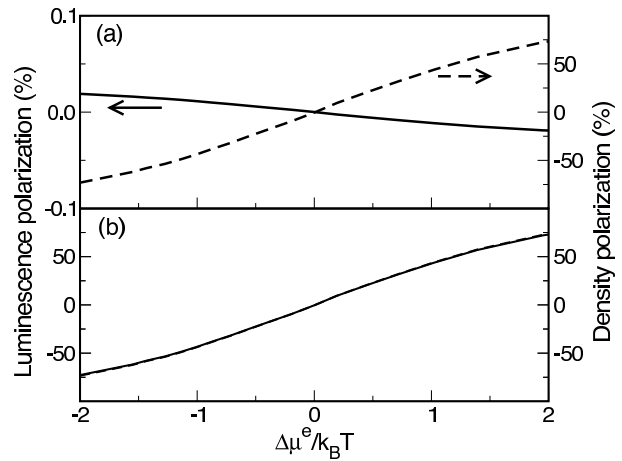}
\caption{Luminescence and density polarizations versus the
Fermi-level splitting for spin-LED structures using spin-polarized
electrons. Panels (a) and (b) are for edge emission and vertical
emission, respectively. The solid lines describe the luminescence
polarization and the dashed lines describe the density
polarization. The electron density is 10$^{15}$ cm$^{-3}$ and the
temperature is 6 K.}\label{elec-emiss}
\end{figure}

The $P_\ell$ and $P_d$ in the QW as a function of the
Fermi-energy splitting for both edge- and vertical-emission
configurations are shown in Fig.~\ref{elec-emiss}. The $P_\ell$ for electron-spin injection is very
weak, at least one order of magnitude smaller than that for hole
spin injection. This is in agreement with recent experiments of
II-VI spin-LED structures using electron spin injection\cite{molenkamp2}, but not with measurements on Zener tunneling diodes\cite{youngsst}. Despite the low $P_\ell$, electrons in the QW can be fully
polarized along the $x$-axis. The large $P_d$ along
the $x$-axis occurs because the spin-up and spin-down states in
the conduction band are still nearly degenerate and a state with
$|c_{\uparrow_x(\downarrow_x)}\rangle$ characteristics can be constructed
from states with $|c_{\uparrow_z(\downarrow_z)}\rangle $ characteristics.
The vertical-emission electron spin LED can have a strong circular
polarization, and $P_\ell\approx P_{d,s}$.

Circularly polarized luminescence has been calculated from
spin LEDs for both hole- and electron-spin injection. When
spin-polarized carriers move from the spacer to the QW, they try
to maintain  their {\it orbital} coherence. Our calculations can
explain many features of the circular polarizations of both edge and vertical
emission observed in Mn-doped ferromagnetic semiconductor spin
LEDs. Our results also indicate that spin-polarized holes give
rise to a much stronger in-plane circular polarization of
luminescence than do spin-polarized electrons, consistent with the
experimental measurements of II-VI spin LEDs using spin polarized
electrons. 

This work was supported by DARPA/ARO DAAD19-01-0490.

\end{document}